\documentstyle[twocolumn]{article}
\def\be{\begin{equation}}
\def\ee{\end{equation}}

\begin{document}

\begin{center}
{\Large {\bf General Relativity Requires Absolute Space and Time}} \\

\vspace{0.5cm}
Rainer W. K\"uhne \\
{\em Lechstr. 63, 38120 Braunschweig, Germany}
\end{center}

\vspace{0.8cm}

\noindent
{\bf We examine two far-reaching and somewhat heretic 
consequences of General Relativity. (i) It requires a cosmology which 
includes a preferred rest frame, absolute space and time. 
(ii) A rotating universe and time travel are strict solutions of General 
Relativity.}

\vspace{0.8cm}

\section{Space and Time Before General Relativity}
\noindent
According to Aristotle, the Earth was resting in the centre of the universe. 
He considered the terrestrial frame as a preferred frame and all motion 
relative to the Earth as absolute motion. Space and time were absolute 
\cite{Aristotle}.

In the days of Galileo the heliocentric model of Copernicus 
\cite{Copernicus} was valid. The Sun was thought to be resting within the 
centre of the universe and defining a preferred frame. Galileo argued that 
only relative motion was observed but not absolute motion. However, to 
fix motion he considered it as necessary to have not only relative motion, 
but also absolute motion \cite{Galileo}.

Newton introduced the mathematical description of Galileo's kinematics. 
His equations described only relative motion. Absolute motion did not 
appear in his equations \cite{Newton}.

This inspired Leibniz to suggest that absolute motion is not required 
by the classical mechanics introduced by Galileo and Newton \cite{Leibniz}. 

Huyghens introduced the wave theory of light. According to his theory, 
light waves propagate via oscillations of a new medium which consists 
of very tiny particles, which he named aether particles. He considered 
the rest frame of the luminiferous aether as a preferred frame 
\cite{Huyghens}.

The aether concept reappeared in Maxwell's theory of classical 
electrodynamics \cite{Maxwell}. Faraday \cite{Faraday} unified Coulomb's 
theory of electricity \cite{Coulomb} with Amp\`ere's theory of magnetism 
\cite{Ampere}. Maxwell unified Faraday's 
theory with Huyghens' wave theory of light, where in Maxwell's theory 
light is considered as an oscillating electromagnetic wave which 
propagates through the luminiferous aether of Huyghens.

We all know that the classical kinematics was replaced by Einstein's 
Special Relativity \cite{SR}. Less known is that Special Relativity is not 
able to answer several problems that were explained by classical mechanics.

According to the relativity principle of Special Relativity, all inertial 
frames are equivalent, there is no preferred frame. Absolute motion is not 
required, only the relative motion between the inertial frames is needed. 
The postulated absence of an absolute frame prohibits the existence of 
an aether \cite{SR}.

According to Special Relativity, each inertial frame has its own relative 
time. One can infer via the 
Lorentz transformations \cite{Lorentz} on the time of the other inertial 
frames. Absolute space and time do not exist. Furthermore, space is 
homogeneous and isotropic, there does not exist any rotational axis of 
the universe.

It is often believed that the Michelson-Morley experiment \cite{Michelson} 
confirmed the relativity principle and refuted the existence of a 
preferred frame. This believe is not correct. In fact, the result of 
the Michelson-Morley experiment disproved the existence of a preferred 
frame only if Galilei invariance is assumed. The experiment can be 
completely explained by using Lorentz invariance alone, the relativity 
principle is not required.

By the way, the relativity principle is not a phenomenon that belongs 
solely to Special Relativity. According to Leibniz it can be applied also 
to classical mechanics.

Einstein's theory of Special Relativity has three problems.

(i) The space of Special Relativity is empty. There are no entities apart 
from the observers and the observed objects in the inertial frames. 
By contrast, the space of classical mechanics can be filled with, say, 
radiation or turbulent fluids.

(ii) Without the concept of an aether Special Relativity can only 
describe but not explain why electric and magnetic fields oscillate in 
propagating light waves.

(iii) Special Relativity does not satisfy the equivalence principle 
\cite{EP} of General Relativity, according to which inertial mass and 
gravitational mass are identical. Special Relativity considers only 
inertial mass.

Special Relativity is a valid approximation of reality which is appropriate 
for the description of most of the physical phenomena examined until 
the beginning of the twenty-first century. However, the macroscopic 
properties of space and time are better described by General Relativity.

\section{General Relativity: Absolute Space and Time}

\noindent
In 1915 Einstein presented the field equations of General Relativity 
\cite{EFE} and in 1916 he presented the first comprehensive article on 
his theory \cite{GR}. In a later work he showed an analogy between 
Maxwell's theory and General Relativity. The solutions of the free 
Maxwell equations are electromagnetic waves while the solutions of the 
free Einstein field equations are gravitational waves which propagate 
on an oscillating metric \cite{grwaves}. As a consequence, Einstein 
called space the aether of General Relativity \cite{aether}. However, 
even within the framework of General Relativity do electromagnetic waves 
not propagate through a luminiferous aether.

Einstein applied the field equations of General Relativity on the entire 
universe \cite{cosmo}. He presented a solution of a homogeneous, 
isotropic, and static universe, where the space has a positive 
curvature. This model became known as the Einstein universe. However, 
de Sitter has shown that the Einstein universe is not stable against 
density fluctuations \cite{desitter}.

This problem was solved by Friedmann and Lema\^itre who suggested a 
homogeneous and isotropic expanding universe where the space is curved 
\cite{Friedmann}.

Robertson and Walker presented a metric for a homogeneous and isotropic 
universe \cite{Robertson}. According to G\"odel this metric requires an 
absolute time \cite{Godel}. In any homogeneous and isotropic cosmology 
the Hubble constant \cite{Hubble} and its inverse, the Hubble age of 
the universe, are absolute and not relative quantities. In the 
Friedmann-Lema\^itre universe there exists a relation between the actual 
age of the universe and the Hubble age.

According to Bondi and Gold, a preferred motion is given at each point 
of space by cosmological observations, namely the redshift-distance 
relation generated by the Hubble effect. It appears isotropic only 
for a unique rest frame \cite{Bondi}.

I argued that the Friedmann-Lema\^itre universe has a finite age and 
therefore a finite light cone. The centre-of-mass frame of this Hubble 
sphere can be regarded as a preferred frame \cite{MMM}.

After the discovery of the cosmic microwave background radiation by 
Penzias and Wilson \cite{Penzias}, it was predicted that it should have 
a dipole anisotropy generated by the Doppler effect by the Earth's 
motion. This dipole anisotropy was predicted in accordance with 
Lorentz invariance \cite{PWBC} and later discovered experimentally 
\cite{Smoot}. Peebles called these experiments ``aether drift 
experiments'' \cite{Peebles}.

The preferred frames defined by the Robertson-Walker metric, the 
Hubble effect, and the cosmic microwave background radiation are 
probably identical. In this case the absolute motion of the Sun was 
determined by the dipole anisotropy experiments of the 
cosmic microwave background radiation to be $(371 \pm 1)$ km/s.

I suggested that this aether drift can give rise to local physical effects. 
I introduced the theory of quantum electromagnetodynamics \cite{MMM}. 
It is a generalization of quantum electrodynamics \cite{QED} which 
includes Dirac's magnetic monopoles \cite{Dirac} and two kinds of photons, 
Einstein's electric photon \cite{photon} and Salam's magnetic photon 
\cite{Salam}. I predicted that every light source which emits electric 
photons does emit also magnetic photons. The ratio between the interaction 
cross-sections of the magnetic photon and the electric photon shall 
depend on the aether drift of the laboratory. The results of recent 
experiments to test my theory may be interpreted as preliminary 
evidence for these magnetic photon rays. These experiments were 
performed in Vienna/Austria by Alipasha Vaziri in February 2002 and 
in Madison/Wisconsin by Roderic Lakes in March and June 2002.

\section{General Relativity: Rotating Universe and Time Travel}

\noindent
It is well-known that planets, stars, and galaxies rotate. So Lanczos 
and Gamow speculated that the entire universe may rotate and that the 
rotating universe might have generated the rotation of the galaxies 
\cite{Gamow}.

G\"odel was the first to show that a rotating universe is a strict 
solution of Einstein's field equations for a homogeneous and 
anisotropic universe. He considered a non-expanding universe and has 
shown that it allows closed time-like curves, i.e. time-travel. 
He predicted that the original order of the rotation axes of galaxies was 
parallel to the universal rotation axis \cite{Godel}.

Raychaudhuri presented a model for an expanding and rotating universe 
which is a generalization of both the Friedmann-Lema\^itre universe 
and the G\"odel universe. This cosmology, too, includes closed 
time-like curves \cite{Raychaudhuri}.

Possibly, the Raychaudhuri universe did not start from a singularity 
(big bang), but from a closed time-like curve, i.e. from a time-machine.

Gregory, Thompson, and Tifft discovered that the distribution of the 
rotation axes for both the spiral and ellipsoid galaxies of the 
filament-like Perseus-Pisces supercluster is bimodal. One of the peaks 
is roughly aligned with the major axis of the supercluster while the 
second peak is roughly $90^{\circ}$ from the first \cite{Gregory}. 
This anisotropic distribution cannot be explained by conventional 
models of galaxy-formation. Therefore I suggested that this might be a 
remnant of the original aligned distribution of galactic rotation axes 
generated by a rotating universe \cite{axis}.

A rotating universe with both vorticity and shear would generate an 
anisotropy of the cosmic microwave background radiation. Collins and 
Hawking were able to set tight bounds on this effect \cite{Hawking}. 
However, Korotky and Obukhov showed that the generation of this anisotropy 
is an effect of shear and not of vorticity alone. So the observed isotropy 
of the cosmic microwave background radiation does not contradict the idea 
of a rotating universe, where the rotation period could be as high as 
the Hubble age of the universe \cite{Obukhov}.

There is some discussion whether General Relativity could allow local 
time-machines. Carter has shown that the Kerr metric \cite{Kerr} of 
rotating spherical bodies can generate closed time-like curves 
\cite{Carter}. This inspired Tipler to investigate a rapidly rotating 
cylinder with 100 km length, 15 km radius, $10^{14}$g/cm$^{3}$ density, 
and a rotational speed of 70\% of the speed of light. This object yielded 
closed time-like curves \cite{Tipler}. However, until now it has not been 
proved that an observer outside the gravitational field would also see 
time-travel.

To conclude, General Relativity requires a cosmology which includes a 
preferred frame, absolute space and time and which may include a 
rotating universe and time-travel. Such a universe may have originated 
not from a singularity (big bang), but from a closed time-like curve 
(time-machine).


\begin{thebibliography}{99}
\bibitem{Aristotle}
Aristotle, De caelo (4th century BC).
\bibitem{Copernicus}
N. Copernicus, De revolutionibus orbium coelestium (1543).
\bibitem{Galileo}
G. Galilei, Discorsi e dimostrazioni matematiche intorno a due nuove 
scienze attenenti alla meccanica ed i movimente locali (Leida, Elsevier, 1638).
\bibitem{Newton}
I. Newton, Philosophiae naturalis principia mathematica (London, 1687).
\bibitem{Leibniz}
G. W. Leibniz, Third letter to S. Clarke (1716).
\bibitem{Huyghens}
C. Huyghens, Trait\'e de la lumi\`ere (1690).
\bibitem{Maxwell}
J. C. Maxwell, A Treatise on Electricity and Magnetism 
(Oxford, Clarendon Press, 1873).
\bibitem{Faraday}
M. Faraday, Experimental Researches in Electricity, Vol. I 
(London, Taylor and Francis, 1839). \\
M. Faraday, Experimental Researches in Electricity, Vol. II 
(London, Richard and John Edward Taylor, 1844). \\
M. Faraday, Experimental Researches in Electricity, Vol. III 
(London, Taylor and Francis, 1855). 
\bibitem{Coulomb}
C. A. Coulomb, {\it Hist. M\'em. l'Acad. R. Sci.}, p. 569 (1785). \\
C. A. Coulomb, {\it Hist. M\'em. l'Acad. R. Sci.}, p. 578 (1785). \\
C. A. Coulomb, {\it Hist. M\'em. l'Acad. R. Sci.}, p. 612 (1785). \\
C. A. Coulomb, {\it Hist. M\'em. l'Acad. R. Sci.}, p. 67 (1786). 
\bibitem{Ampere}
A.-M. Amp\`ere, {\it Ann. Chim. Phys.} {\bf 15}, 59 (1820). \\
A.-M. Amp\`ere, {\it Ann. Chim. Phys.} {\bf 15}, 170 (1820).
\bibitem{SR}
A. Einstein, {\it Ann. Phys. (Leipzig)} {\bf 17}, 891 (1905).
\bibitem{Lorentz}
J. Larmor, Aether and Matter (Cambridge, University Press, 1900). \\
H. A. Lorentz, {\it Proc. R. Acad. Amsterdam} {\bf 6}, 809 (1904).
\bibitem{Michelson}
A. A. Michelson, {\it Am. J. Sci.} {\bf 22}, 120 (1881). \\
A. A. Michelson and E. W. Morley, {\it Am. J. Sci.} {\bf 34}, 333 (1887).
\bibitem{EP}
A. Einstein, {\it Ann. Phys. (Leipzig)} {\bf 38}, 355 (1912).
\bibitem{EFE}
A. Einstein, {\it S.-B. Preuss. Akad. Wiss.}, p. 844 (1915).
\bibitem{GR}
A. Einstein, {\it Ann. Phys. (Leipzig)} {\bf 49}, 769 (1916).
\bibitem{grwaves}
A. Einstein, {\it S.-B. Preuss. Akad. Wiss.}, p. 154 (1918).
\bibitem{aether}
A. Einstein, \"Ather und Relativit\"atstheorie 
(Berlin, Springer-Verlag, 1920).
\bibitem{cosmo}
A. Einstein, {\it S.-B. Preuss. Akad. Wiss.}, p. 142 (1917).
\bibitem{desitter}
W. de Sitter, {\it Konin. Ned. Akad. Wetenschappen} {\bf 19}, 1217 (1917).
\bibitem{Friedmann}
A. Friedmann, {\it Z. Phys.} {\bf 10}, 377 (1922). \\
A. Friedmann, {\it Z. Phys.} {\bf 21}, 326 (1924). \\
G. Lema\^itre, {\it Ann. Soc. Sci. Brux.} {\bf 47}, 49 (1927).
\bibitem{Robertson}
H. P. Robertson, {\it Astrophys. J.} {\bf 82}, 284 (1935). \\
A. G. Walker, {\it Proc. London Math. Soc.} {\bf 42}, 90 (1936).
\bibitem{Godel}
K. G\"odel, {\it Rev. Mod. Phys.} {\bf 21}, 447 (1949).
\bibitem{Hubble}
E. P. Hubble, {\it Proc. Nat. Acad. Sci.} {\bf 15}, 168 (1929).
\bibitem{Bondi}
H. Bondi and T. Gold, {\it Nature} {\bf 169}, 146 (1952).
\bibitem{MMM}
R. W. K\"uhne, {\it Mod. Phys. Lett. A} {\bf 12}, 3153 (1997).
\bibitem{Penzias}
A. A. Penzias and R. W. Wilson, {\it Astrophys. J.} {\bf 142}, 419 (1965).
\bibitem{PWBC}
P. J. E. Peebles and D. T. Wilkinson, {\it Phys. Rev.} {\bf 174}, 2168 
(1968). \\
R. N. Bracewell and E. K. Conklin, {\it Nature} {\bf 219}, 1343 (1968).
\bibitem{Smoot}
G. F. Smoot, M. V. Gorenstein, and R. A. Muller, 
{\it Phys. Rev. Lett.} {\bf 39}, 898 (1977).
\bibitem{Peebles}
P. J. E. Peebles, Physical Cosmology (Princeton, University Press, 1971).
\bibitem{QED}
S. Tomonaga, {\it Phys. Rev.} {\bf 74}, 224 (1948). \\
J. Schwinger, {\it Phys. Rev.} {\bf 74}, 1439 (1948). \\
J. Schwinger, {\it Phys. Rev.} {\bf 75}, 651 (1949). \\
J. Schwinger, {\it Phys. Rev.} {\bf 76}, 790 (1949). \\
R. P. Feynman, {\it Phys. Rev.} {\bf 76}, 749 (1949). \\
R. P. Feynman, {\it Phys. Rev.} {\bf 76}, 769 (1949). \\
F. J. Dyson, {\it Phys. Rev.} {\bf 75}, 486 (1949). \\
F. J. Dyson, {\it Phys. Rev.} {\bf 75}, 1736 (1949). 
\bibitem{Dirac}
P. A. M. Dirac, {\it Proc. R. Soc. A} {\bf 133}, 60 (1931).
\bibitem{photon}
A. Einstein, {\it Ann. Phys. (Leipzig)} {\bf 17}, 132 (1905).
\bibitem{Salam}
A. Salam, {\it Phys. Lett.} {\bf 22}, 683 (1966).
\bibitem{Gamow}
C. Lanczos, {\it Z. Phys.} {\bf 21}, 73 (1924). \\
G. Gamow, {\it Nature} {\bf 158}, 549 (1946).
\bibitem{Raychaudhuri}
A. Raychaudhuri, {\it Phys. Rev.} {\bf 98}, 1123 (1955).
\bibitem{Gregory}
S. A. Gregory, L. A. Thompson, and W. G. Tifft, {\it Astrophys. J.} 
{\bf 243}, 411 (1981).
\bibitem{axis}
R. W. K\"uhne, {\it Mod. Phys. Lett. A} {\bf 12}, 2473 (1997).
\bibitem{Hawking}
S. W. Hawking, {\it Mon. Not. R. Astron. Soc.} {\bf 142}, 129 (1969). \\
C. B. Collins and S. W. Hawking, {\it Mon. Not. R. Astron. Soc.} 
{\bf 162}, 307 (1973).
\bibitem{Obukhov}
V. A. Korotky and Yu. N. Obukhov, {\it Sov. Phys. JETP} {\bf 72}, 11 
(1991). \\
Yu. N. Obukhov, {\it Gen. Relativ. Grav.} {\bf 24}, 121 (1992). \\
S. Carneiro and G. A. Mena Marug\'an, {\it Phys. Rev. D} {\bf 64}, 
083502 (2001).
\bibitem{Kerr}
R. P. Kerr, {\it Phys. Rev. Lett.} {\bf 11}, 237 (1963). 
\bibitem{Carter}
B. Carter, {\it Phys. Rev.} {\bf 174}, 1559 (1968).
\bibitem{Tipler}
F. J. Tipler, {\it Phys. Rev. D} {\bf 9}, 2203 (1974).

\end{thebibliography}
\end{document}